\documentclass[aps,twocolumn,superscriptaddress]{revtex4}

\usepackage{amsmath,amssymb}
\usepackage{epsfig}
\usepackage{graphicx}
\begin{document}

\title{
Late time solutions for inhomogeneous $\Lambda$CDM cosmology, their characterization and observation
}

\author{Toby Wiseman and Benjamin Withers}
\affiliation{Theoretical Physics Group, Blackett Laboratory, Imperial College, London SW7 2AZ, U.K.}

\date{May 2010}

\begin{abstract}

Assuming homogeneous isotropic $\Lambda$CDM cosmology allows $\Lambda$, spatial curvature and dark matter density to be inferred from large scale structure observations such as supernovae. The purpose of this paper is to extend this to allow observations to measure or constrain inhomogeneity and anisotropy.
We obtain the general inhomogeneous anisotropic $\Lambda$CDM solution which is locally asymptotic to an expanding de Sitter solution as a late time expansion using Starobinsky's method (analogous to the `holographic renormalization' technique in AdS/CFT) together with a resummation of the expansion. The dark matter is modeled as perfect dust fluid. 
The terms in the expansion systematically describe inhomogeneous and anisotropic deformations of an expanding FLRW solution, and are given as a spatial derivative expansion in terms of data characterizing the solution - a 3-metric and a perturbation of that 3-metric.
Leading terms describe inhomogeneity and anisotropy on the scale set by the cosmological constant, approximately the horizon scale today. Higher terms in the expansion describe shorter scale variations.
We compute the luminosity distance-redshift relation and argue that
comparison with current and future observation would allow a partial reconstruction of the characterizing data.
We also comment on smoothing these solutions noting that geometric flows (such as Ricci flow) applied to the characterizing data provide a canonical averaging method.

\end{abstract}


\maketitle

\section{Introduction}

Modern large scale structure observations, such as the SDSS low redshift studies of `standard candle' supernovae \cite{Frieman:2007mr,Sako:2007ms}, afford us the capability to directly test our cosmological model assuming only the Einstein equations, matter content relevant for our current epoch, and homogeneity and isotropy (and hence the FLRW metric). Historically Hubble's measurement of the current expansion rate was the first test of this type. Recently the famous results of the Supernova Search Team \cite{Riess:1998cb} and the Supernova Cosmology Project \cite{Perlmutter:1998np} allowed comparison of luminosity distance-redshift for various matter content models with supernovae data up to redshift $\sim 5$, showing the likely current domination of a positive cosmological constant over dark matter.

An important point is that unlike CMB observations, large scale structure allows inferences to be made about the parameters and dynamics of a model without any assumption on initial conditions, except for the, albeit strong, assumptions of homogeneity and isotropy. While the cosmological constant can be estimated from CMB measurements, this inference is far less direct, and is subject to a choice of inflationary initial conditions. Here we are certainly not arguing that one should not take inflationary initial conditions seriously, but rather we wish to emphasize that when one makes inferences about physical models from observational data, the strength of the parameter estimation implicitly depends on the strength of the assumptions made. It is useful to have a variety of methods to allow estimation with differing assumptions.

It is worth emphasizing that in the context of inflation it is the inflationary mechanism that suppresses large scale inhomogeneity, anisotropy as well as spatial curvature. Thus if one is interested in assessing the magnitude of spatial curvature by measurement from observation, for example if one thought inflation lasted only a minimal time and therefore interesting structure may be seen on large scales (e.g. \cite{Efstathiou:2003hk}), it is conceptually inconsistent to include spatial curvature. One should consider general anisotropic inhomogeneous deformations at the same time as introducing spatial curvature, unless one supplements the initial conditions before inflation with homogeneity and isotropy, as for example occurs after semiclassical bubble nucleation of an open universe \cite{Coleman:1980aw}. 

The purpose of this paper is to allow the comparison of a $\Lambda$CDM cosmology with large scale structure data such as supernovae, \emph{without making the assumptions of homogeneity or isotropy} on large scales. This is not because we are arguing the universe is far from homogeneous or isotropic, but simply because in order to argue the universe is homogeneous and isotropic, ideally one should deduce this from the data without assuming it. Furthermore with future measurements one would hope to be able to quantify precisely \emph{how far} from homogeneity and isotropy the universe is, or at least put rigorous upper bounds on how large these quantities are on different scales.

Of course in order to do any of this, one must be able to compute and characterize general inhomogeneous solutions for a $\Lambda$CDM cosmology. In addition, one must then be able to compute quantities relevant for observation, such as the luminosity distance-redshift relation, and show how this allows inferences about the data that characterizes the solutions to be made from real large scale structure data.
Certain exact solutions have allowed such analysis to be performed for very special cases, for instance using the LTB solution which is isotropic but inhomogeneous, or the Bianchi models which are anisotropic but homogeneous \cite{Lemaitre:1933gd,Tolman:1934za,Bondi:1947av,Mustapha:1997xb,Mustapha:1998jb,Celerier:1999hp,Chung:2006xh}, but general exact solutions are likely not to exist in any closed form.

In this paper we show how to construct general inhomogeneous, anisotropic solutions of the Einstein equations with cosmological constant and dust fluid as a model for baryonic and dark matter using a method which we call a `late time expansion'. This method was first developed by Starobinsky \cite{Starobinsky:1982mr} in the context of inflation, and is similar to methods employed in AdS/CFT holographic renormalization (see \cite{Skenderis:2002wp}). 
Rather than starting from the conventional approach of cosmological perturbation theory about the FLRW solution, instead we use the fact that provided the local universe is dynamically dominated by the cosmological constant, one can describe the solution in a precise way as a deformation about de Sitter space, characterized by its late time asymptotic behaviour, or what we shall term the \emph{final} boundary conditions. The leading late time behaviour of the solution is a universe that is locally de Sitter. Choosing the late time de Sitter conformal boundary to have proper time, $t \rightarrow \infty$, the full solution then takes the form of an expansion in positive powers of $e^{- \Lambda t/3}$, where higher terms in the expansion give inhomogeneous corrections to this local de Sitter behaviour, and are given in terms of increasing numbers of derivatives acting on the final boundary data. We find the form of this final data is given by a 3-metric and a perturbation of that 3-metric. The leading terms in the late time expansion describe inhomogeneity and anisotropy on the scale set by the cosmological constant, and hence for our universe would correspond to approximately today's horizon size deformations. Higher terms in the series then govern derivative corrections which correspond to increasingly smaller scale inhomogeneity and anisotropy. 

The first part of our paper is devoted to giving the solution for $\Lambda$CDM, taking dust fluid, using the `late time expansion' about an asymptotically locally de Sitter solution with flat spatial slices. We give results beyond leading order, and in particular discuss the final data that arises in the expansion and characterizes the solution. For dust fluid it is possible to take a normal coordinate system adapted to the conformal boundary that is comoving with the dust - this is not possible for fluids with pressure. This simplifies the calculation considerably, and is what allows us to give results straightforwardly beyond the leading order. 

In the second part of the paper, having showed how to give inhomogeneous anisotropic solutions as a late time expansion deforming about flat sliced de Sitter, we go on to consider the convergence of the late time expansion, arguing that whilst one might naively expect it to break down around $\Lambda$-matter equality, it actually is far better, extending back to very early epochs. We then discuss the drawback of our first expansion, namely that homogeneous isotropic dust and curvature are also described as deformations and treated order by order in the late time expansion, even though we know the exact homogeneous isotropic solution is an expanding FRLW universe, whose solution we may treat analytically. However, we show that the result may simply be modified to give an expansion about the exact expanding FRLW solution using a simple but powerful series resummation on the late time expansion. The terms in the expansion then describe deformations only involving inhomogeneity and isotropy. If the deviations about FLRW solution are small we believe, although here do not prove, that this late time expansion converges back to high redshifts, certainly sufficient to describe the region relevant for large scale structure observations.

The third part of the paper is then concerned with applying these inhomogeneous and anisotropic solutions to observations of our universe. We compute the luminosity distance-redshift relation explicitly for the first two non-trivial orders in the expansion. We then discuss how one might in principle use this to compare to supernovae data and therefore extract information about the characterizing data, the boundary 3-metric and its perturbation. In particular we discuss how the boundary data is extracted as a derivative expansion about the point corresponding to the observers asymptotic position, and that fitting increasingly high orders in the luminosity distance-redshift relation expansion to data gives information about increasingly high derivatives of the local curvature of the 3-metric and perturbation. Thus information characterizing the inhomogeneous and anisotropic universe we live in can be systematically extracted from observational data, provided that, as we argue, we are living in and observing the regime where the late time expansion converges. In fact not all characterizing data can be extracted from this luminosity distance-redshift measurement, and we deduce the subset of the data that may be extracted. 

We briefly conclude with a discussion of averaging or smoothing of our inhomogeneous solutions. Since in our method the late time inhomogeneous cosmologies are determined by geometric final data consisting of a 3-metric and its perturbation, geometric flows, such as Ricci flow, provide the canonical notion of averaging this characterizing data.
Considering this we arrive at a natural averaging of our solutions similar in spirit although rather different in origin to that discussed by Buchert and Carfora \cite{Buchert:2002ht,Buchert:2003}.  It is simple to argue that provided one begins with small deformations of FLRW, such smoothing flows to the FLRW solution.

\section{Late time expansion about de Sitter}

Following the method of Starobinsky's late time expansion \cite{Starobinsky:1982mr}, utilized in the context of inflation, we now show how to construct the general cosmological solution in a $\Lambda$CDM model that is locally asymptotic to de Sitter, taking the dark matter and baryonic matter to be modeled by a single dust fluid. 

Unlike a conventional treatment of inhomogeneity as a deformation about de Sitter in a perturbative expansion in deformation amplitude, here Starobinsky's method treats the solution as a deformation about de Sitter as a late time expansion, where the perturbation parameter instead measures how late the region being described is. No assumption of homogeneity is necessary, and inhomogeneity is treated fully non-perturbatively.

A key technical tool that we may use is to choose a conformal frame where the dust is comoving. In fact this is possible only for dust fluid, and very much simplifies the construction of the solution, which previously was given by Starobinsky and later more rigourously by Rendall \cite{Rendall:2003ks} for cosmological constant and general perfect fluid but only to leading order. Here, for dust, we are able to straightforwardly derive several orders in the expansion.

We proceed by writing the metric covering the late time region of interest in a Gaussian normal form to constant $y$ surfaces,
\begin{eqnarray}
\label{eq:metric1}
ds^2 =  \frac{1}{y^2} \left( - \frac{dy^2}{H^2}  + g_{ij}(x,y) dx^i dx^j \right)
\end{eqnarray}
where, as before, $x^i$ for $i=1,2,3$ give spatial coordinates on the constant $y$ surface and $H^2 = \Lambda/3$, and would give the Hubble rate for the de Sitter geometry corresponding to the cosmological constant $\Lambda$. By construction of normal coordinates the vector $\partial/\partial y$ is tangent to timelike geodesics normal to the constant $y$ surfaces. We have the freedom to choose these constant $y$ surfaces to comove with the dust fluid so that the stress tensor takes the simple form,
\begin{eqnarray}
T_{yy} = \frac{1}{H^2 y^2} \left( \Lambda + \rho \right) \; , \quad T_{yi} = 0 \; , \quad T_{ij} =  - \frac{\Lambda}{y^2} g_{ij}
\end{eqnarray}
and this fixes our choice of conformal frame. Taking a different frame would require us to introduce a velocity field for the dust fluid.

We decompose the Einstein equations (taking $8 \pi G = 1$) to give a scalar and vector equation with respect to the 3-metric $g_{ij}$,
\begin{eqnarray}
\ddot{g} - \frac{1}{y} \dot{g} - \frac{1}{2} \dot{g}_{ij} \dot{g}^{ij} &=& - \frac{1}{H^2 y^2} \rho \nonumber \\
\nabla^j \dot{g}_{ij} - \nabla_i \dot{g} &=& 0
\end{eqnarray}
and in addition a tensor equation,
\begin{eqnarray}
 \ddot{g}_{ij}  - \frac{2}{y} \dot{g}_{ij}  + \frac{1}{8} g_{ij} \left( \dot{g}_{mn}\dot{g}^{mn}- \dot{g}^2 \right) -  \dot{g}_{im} \dot{g}_{j}^{~m}   && \nonumber \\
 + \frac{1}{2} \dot{g} \, \dot{g}_{ij} + \frac{2}{H^2} \left( R_{ij} - \frac{1}{4} g_{ij} R \right) &=& 0
 \label{eq:tensoreq}
\end{eqnarray}
where indices are raised and lowered with respect to $g_{ij}$, $\nabla_i$ is the Levi-Civita connection of $g_{ij}$, $R_{ij}$ is its Ricci tensor, and we define $\dot{g}_{ij} = \partial_y g_{ij}$, $\dot{g} = g^{ij} \dot{g}_{ij}$ and likewise $\ddot{g}_{ij} = \partial^2_y g_{ij}$, $\ddot{g} = g^{ij} \ddot{g}_{ij}$. As usual, the fluid dust equation of motion is implied by the Einstein equations due to the contracted Bianchi identity.

We now consider inhomogeneous solutions and use a late time expansion as a tool to solve the equation. Firstly we consider solving the equation \eqref{eq:tensoreq} as an expansion in $y$. One obtains,
\begin{eqnarray}
\label{eq:soln}
g_{ij}(y,x) &=& \bar{g}_{ij}(x) + y^2 a^{(0)}_{ij}(x) + y^3 h_{ij}(x)  \nonumber \\
&& \quad + y^4 a^{(1)}_{ij}(x) + y^5 a^{(2)}_{ij}(x) + \ldots
\end{eqnarray}
where the two pieces of data for the solution of this second order equation are provided by the symmetric $\bar{g}_{ij}(x)$ and $h_{ij}(x)$. 
We may interpret $\bar{g}_{ij}(x)$ as the metric of a 3-geometry -- the conformal boundary metric of the full spacetime -- and then the other data $h_{ij}(x)$ we regard as a symmetric tensor living on this geometry.
The terms $a^{(i)}_{ij}$ and all higher terms in the expansion are determined in terms of this metric and the data $h_{ij}$. Explicitly we find these lower order terms to be,
\begin{eqnarray}
a^{(0)}_{ij} & = & \frac{1}{H^2} \left( \bar{R}_{ij} - \frac{1}{4} \bar{g}_{ij} \bar{R} \right) \nonumber \\
a^{(1)}_{ij} & = & a^{(0)}_{im}a^{(0)m}_{~j} - \frac{1}{2} a^{(0)} a^{(0)}_{ij} - \frac{1}{8} \bar{g}_{ij} \left( a^{(0)mn} a^{(0)}_{mn} - (a^{(0)})^2 \right) \nonumber \\
&& \quad  - \frac{1}{2 H^2} \mathcal{O}_{ij}^{~~mn} a^{(0)}_{mn} \nonumber \\
a^{(2)}_{ij} & = & \frac{3}{5} \left( a^{(0)}_{im} h_{j}^{~m} + a^{(0)}_{jm} h_{i}^{~m} \right) - \frac{3}{10} \left( h a^{(0)}_{ij} + a^{(0)} h_{ij} \right) \nonumber \\
&& \quad  -\frac{3}{20} \bar{g}_{ij} \left( a^{(0)}_{mn} h^{mn} - a^{(0)} h \right) - \frac{1}{5 H^2} \mathcal{O}_{ij}^{~~mn} h_{mn}  
\end{eqnarray} 
where $\bar{R}_{ij}$ is the Ricci tensor of the data $\bar{g}_{ij}$ and indices in the terms $a^{(i)}_{ij}$ are lowered and raised with respect to $\bar{g}_{ij}$, and $\bar{\nabla}_i$ is its Levi-Civita connection, and $\mathcal{O}$ is a differential operator that we give below in equation \eqref{eq:Oop}.

We now turn to the remaining scalar and vector equations. The scalar equation simply determines the behaviour of the dust fluid density as,
\begin{eqnarray}
\frac{\rho}{H^2} &=&  - 3 y^3 h  + \frac{3}{2} y^5  a^{(0)} h   + O(y^6)
\end{eqnarray}
The vector equation is non-trivial and constrains the space of solutions. Let us define the vector $\Phi_i \equiv \bar{\nabla}^j \dot{g}_{ij} - \bar{\nabla}_i \dot{g}$, and then assuming the tensor equation \eqref{eq:tensoreq} is satisfied we find,
$\partial_y \Phi_i = \left( \frac{2}{y} - \frac{1}{2} \dot{g} \right) \Phi_i$
which may be solved for each component of $\Phi_i$ separately by quadrature to give, 
\begin{eqnarray}
\Phi_i(y,x) = A_i(x) \, y^2  e^{-\frac{1}{2} \int_0^y d\tilde{y} \, \dot{g}(\tilde{y},x)}
\end{eqnarray}
with integration constants $A_i(x)$. However, evaluating $\Phi_i$ on our late time expansion above, we find, 
\begin{eqnarray}
\Phi_i(y,x) = 3 \left( \bar{ \nabla}^j h_{ij} - \bar{\nabla}_i h  \right) y^2  + O(y^3)
\end{eqnarray}
and hence we obtain $A_i(x) = 3 \left( \bar{ \nabla}^j h_{ij} - \bar{\nabla}_i h  \right)$. Thus in order to set the vector equation to zero, we require that $A_i(x) = 0$, and hence we require a constraint on our initial data $h_{ij}$, $\bar{ \nabla}^j h_{ij} - \bar{\nabla}_i h = 0$.

The form of the operator $\mathcal{O}$, which we previously postponed giving, is, 
\begin{eqnarray}
\label{eq:Oop}
\mathcal{O}_{ij}^{~~mn} T_{mn} &=& - \frac{1}{2} \bar{\nabla}^2 T_{ij} + \frac{1}{2} \bar{\nabla}_i \partial_j T 
- \bar{R}_{i~j}^{~m~n} T_{mn} 
\nonumber \\
&& + \bar{R}_{m(i} T_{j)}^{~~m} - \frac{1}{4} \bar{R} \, T_{ij} + \frac{1}{4} \bar{g}_{ij}  \bar{R}_{mn} T^{mn} \qquad
\end{eqnarray}
for action on a symmetric tensor $T_{ij}$ which obeys $\bar{ \nabla}^j T_{ij} - \bar{\nabla}_i T = 0$ as do both $a^{(0)}_{mn}$ and $h_{mn}$. We note the fact that in 3-dimensions the Riemann tensor which enters the above equation is simply determined by the Ricci tensor as,
\begin{eqnarray}
\label{eq:Riem}
\bar{R}_{ijkl} = 2 \left( \bar{g}_{i[k} \, \bar{R}_{l]j} - \bar{g}_{j[k} \, \bar{R}_{l]i} \right) - \bar{R} \, \bar{g}_{i[k} \, \bar{g}_{l]j}
\end{eqnarray}
due to the vanishing of the Weyl tensor.

This `late time expansion' method was first used by Starobinsky in the context of inflation to argue that inhomogeneity is washed out during inflation. In \cite{Starobinsky:1982mr} Starobinsky stated the result for cosmological constant and general fluid matter, but only to leading order. Recently Rendall \cite{Rendall:2003ks} has given a more rigorous derivation for Starobinsky's somewhat briefly justified leading order results. To our knowledge, Starobinsky's work has not been extended beyond leading order for fluids, and has never been applied to the late time universe, but only to inflation. 

We note that Starobinsky's method is quite analogous to the techniques developed in geometry for hyperbolic space \cite{Fefferman} which are ubiquitous in string theory holography (for a recent review see \cite{Skenderis:2002wp}). Note that in \cite{Gibbons, Skenderis:2002wp} beyond leading order results are given for pure cosmological constant matter. There has been considerable interest in using string theory inspired holographic ideas to give new quantum gravitational theories of the early universe and inflation \cite{Strominger:2001pn,Larsen:2002et,McFadden:2009fg,McFadden:2010na}. We emphasize that whilst some technical methods of our current work bear similarity with those used in holography conceptually our work is very different in character.  We are here concerned only with describing the late time universe and not its early stages, and have no dual holographic description in mind.

We also note that the late time expansion method may apply more generally than to cosmologies that are asymptotically locally de Sitter. For example, in \cite{Barrow:2010wh} analogous expansions have been proposed for asymptotics with future singularities. One might hope that one can also generalize this method to allow a late time expansion for a cosmology with a quintessence field and dust, and this would be an interesting direction for future work.

\subsection{Characterizing solutions}

Our choice of conformal frame makes it straightforward to identify that our late time solution is given by the expansion \eqref{eq:soln} where the data is the (Euclidean signature) 3-geometry, specified by the metric $\bar{g}_{ij}$, together with a symmetric tensor $h_{ij}$ defined on this 3-geometry obeying,
\begin{eqnarray}
\label{eq:constraint}
\bar{ \nabla}^j h_{ij} - \bar{\nabla}_i h = 0 \;. 
\end{eqnarray}
The late time density of the dust is given in terms of the trace, $\rho(y,x) = -3 H h(x) y^3 + O(y^4)$, and hence we require this trace  to be negative definite $h = \bar{g}^{ij} h_{ij} < 0$ to ensure a physical dust density. We note that there is a global scaling invariance of the metric \eqref{eq:metric1},
\begin{eqnarray}
\label{eq:scale}
y \rightarrow y / \lambda \, , \quad
x^i \rightarrow x^i / \lambda \, , \quad
\bar{g}_{ij} \rightarrow \bar{g}_{ij} \, , \quad
h_{ij} \rightarrow \lambda^3 \, h_{ij}
\end{eqnarray}
for $\lambda \in \mathbb{R}^+$. Defining the equivalence, $h_{ij} \sim \lambda^3 h_{ij}$ for $\lambda \in \mathbb{R}^+$, we may then take the data characterizing the solution to be given by the boundary metric and equivalence class of $h_{ij}$, namely $(\bar{g}_{ij}, [ h_{ij} ] )$.

Consider a 3-geometry with metric $\bar{g}_{ij}$ and an infinitesimally perturbed geometry, $\bar{g}_{ij} + \epsilon \, \delta \bar{g}_{ij}$, for infinitesimal $\epsilon$. Due to the ability to rescale the infinitesimal $\epsilon$, we take a geometry and infinitesimal perturbation to be characterized by the pair $(\bar{g}_{ij}, [ \delta \bar{g}_{ij} ])$. Define the vector field,
\begin{eqnarray}
\label{eq:uvec}
u_i \equiv \bar{\nabla}^j \delta \bar{g}_{ij} - \frac{1}{2} \bar{\nabla}_{i} \delta g \; .
\end{eqnarray}
Consider a coordinate system such that the `harmonic gauge' condition $u_i = 0$ holds. This locally fixes all the coordinate freedom. We do not address whether there are global obstructions to the existence of such a gauge, but believe it is plausible that such a gauge always exists and is unique, and assume so from now on.\footnote{We note that the existence of such a harmonic gauge for any perturbation requires the existence of solutions to the elliptic vector equation, $\bar{\nabla}^j \bar{\nabla}_{j} \xi_{i} + \bar{R}_{ij} \xi^{j} = H_i$ for some source vector $H^i$.} 
In such a gauge, taking $h_{ij} = \delta \bar{g}_{ij} - \frac{1}{4} \bar{g}_{ij} \delta \bar{g}$, then $h_{ij}$ obeys precisely the appropriate condition \eqref{eq:constraint}. The physical dust density constraint, $h < 0$, then becomes $\delta g < 0$. Since $\delta \left( \sqrt{\det{\bar{g}_{ij}}} \right) = \frac{1}{2} \sqrt{\det{\bar{g}_{ij}}} \delta g$, this implies the perturbation in harmonic gauge locally decreases 3-volume. 

Thus we conclude that a 3-geometry and perturbation, $(\bar{g}_{ij}, [\delta \bar{g}_{ij}])$, can be mapped to the data $(\bar{g}_{ij}, [h_{ij}])$ of our late time solution, provided the perturbation can be cast in harmonic gauge $u_i = 0$. A physical perturbation in that gauge locally decreases volume, which ensures positive dust density. This 3-geometry and perturbation provide an elegant geometric way to characterize our late time inhomogeneous solutions. 

We note that if one wished to include other matter in addition to the dust fluid, then one would have to introduce additional tensor fields to describe its data. As we know from discussion of holographic renormalization in AdS/CFT, extension to include other matter fields would likely be a straightforward, but messy exercise. We will not be concerned with that here, as from a practical point of view we wish to use this solution to better understand our late time universe, and other forms of matter (radiation, inflaton etc) play little role at late times.

\subsection{Remarks on the late time expansion}

We have seen that the terms in our late time expansion give a derivative expansion in the boundary data $( \bar{g}_{ij}, h_{ij} )$. The derivative expansion is controlled by the cosmological constant scale, $1 / H$. Hence the leading terms in the expansion describe inhomogeneity and anisotropy on the largest scales, namely the late time horizon size. Higher terms in the expansion then describe shorter scales, and in principle the derivative expansion may describe all scales, although this is contingent on the late time expansion converging. In general, one might expect that as one moves to earlier times so that the inhomogeneity becomes significant on scales smaller than the comoving horizon size one would expect convergence to worsen and at some point break down.

\begin{figure*}[]
\includegraphics[width=0.8\textwidth]{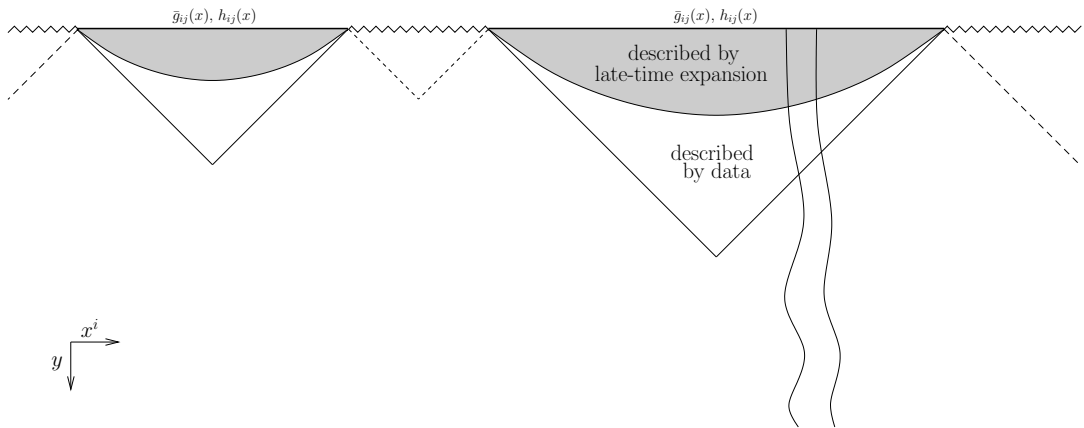}
\caption{\label{latetime}A cartoon of the regions where the late time expansion might converge for a cosmology with both de Sitter and black hole future conformal boundary asymptotics.}
\end{figure*}

The expansion need not describe the global future of the spacetime, but may just describe a local region of the future conformal boundary that is asymptotic to de Sitter. Other behaviours are possible in the future, such as dust or curvature collapse, corresponding to a future conformal boundary that is a black hole singularity. Then the hyperbolic nature of the Einstein equations presumably implies that the late time expansion converges only within some portion of the past light cone of a region that is asymptotic to de Sitter in the future, as illustrated in figure \ref{latetime}. However, we note that for our universe, on large scales, we do expect that the universe is asymptotic to de Sitter, provided that the matter content is a cosmological constant, and dark and baryonic matter is effectively describable by dust fluid. While locally matter may collapse to form black holes, one should average over such small scale inhomogeneity when discussing inhomogeneity with size comparable to today's horizon size as our late time expansion does.

One might wonder what the essential difference is between our solution and a Taylor expansion of data about any Cauchy surface. In effect one can view our solution as a Frobenius expansion about a regular singular point in the Einstein equations, that given by $y \rightarrow 0$.  It is worth emphasizing the following points.
\begin{enumerate}
\item Whilst one can indeed characterize a solution by analytic data on any Cauchy surface and then give the analytic expansion governing its evolution (in the sense of Cauchy-Kowaleski) there is no canonical surface to perform this about. In our construction the future conformal boundary provides a canonical surface. In particular this means that through making observations one can in principle measure model parameters which are independent of the observer.
\item Data to the future of a region of a Cauchy surface is not guaranteed to enter a cosmological constant dominated future -- for example, dust or curvature may come to dominate and lead to a singularity, as illustrated in figure \ref{latetime}. By characterizing our solution with data that lives at the future de Sitter conformal boundary, we have focused by construction precisely only on the regions of cosmology where the cosmological constant comes to dominant, as we as observers see today.
\item On a more technical level, the late time expansion exhibits a separation of the orders at which various matter/curvature contributions first appear. For example, the spatial curvature of the boundary metric first enters at order $y^2$. Dust first enters the expansion at order $y^3$. Had we included radiation, it would not enter until order $y^4$.
This is generally not the case for a Taylor expansion about some Cauchy surface, where all forms of matter and curvature enter at the first order in the expansion. 
We believe that this clean initial separation at low orders in $y$ will prevent possible ambiguity when fitting data, particularly at low redshifts.
\end{enumerate}

We now briefly discuss our expansion from the perspective of AdS/CFT holographic renormalization. If a dual QFT existed, it would live on a 3-d spacetime with metric $\bar{g}_{ij}$. The tensor $h_{ij}$ would be interpreted as giving the boundary theory conserved stress tensor as $T_{ij} = h_{ij} - \bar{g}_{ij} h$. Due to the presence of the dust matter, the stress tensor is not trace free, and hence the dual theory is not conformal. The dust would be associated to a marginal dimension 3 operator in the dual theory,  and a trace perturbation of $\bar{g}_{ij}$ and $h_{ij}$ give the non-normalizable and normalizable 
gravity modes respectively that correspond to the source and vev of this operator. An interesting point is that there are no irrelevant operators, only the marginal graviton and dust. The absence of irrelevant operators corresponds in the gravity to our late time expansion being a stable attractor solution with no perturbative unstable modes that might destroy the de Sitter late time asymptotics.

\subsection{Convergence and application to our universe}

Using the late time expansion method we have shown how to construct the general inhomogeneous anisotropic late time solution as a deformation of de Sitter. Then one expects that provided one lives at late times, so that the cosmological constant dominates, then the expansion would converge, and one could truncate the expansion to obtain a controlled approximation to the true solution. From the practical perspective of application to our universe a natural question follows. Do we live late enough for the series to converge? Recall that in our expansion above the homogeneous dust component is also treated as part of the deformation about de Sitter. Whilst we expect $\Lambda$ to dominate matter today (as $\Omega_m \sim 0.3$), it is not parametrically larger. Naively one might worry even whether the series converges restricting to FLRW for us today, or certainly back to earlier redshifts, say $z \sim 5$, where supernovae yield important large scale structure observations. At such an epoch matter dominates $\Lambda$ for FLRW and thus one would naively assume this is too early an epoch for our late time expansion to describe.

This reasoning however is too quick as we shall now show. Rather surprisingly, for FLRW relevant for our universe, the expansion converges back to the big bang (in this dust-$\Lambda$ model). Let us consider a flat sliced FLRW universe. The solution is simply written using our form of the metric as,
\begin{eqnarray}
\label{eq:dustexp}
g_{ij} &=& \left( 1 - \tfrac{\rho_0}{12 H^2} y^3 \right)^{4/3} \delta_{ij} \nonumber \\
&\simeq& \left( 1 - \tfrac{\rho_0}{9 H^2} y^3 + \tfrac{\rho_0^2}{648 H^4} y^6 + \tfrac{\rho_0^3}{34992 H^6} y^9 + \ldots \right) \delta_{ij} \qquad
\end{eqnarray}
so that $\rho \sim \rho_0 y^3 + \ldots$. One can see from the analytic form of the exact solution that convergence of the series in fact will extend back to the big bang time, as this represents the only non-analytic behaviour (a branch point) in the solution. To approximate the solution well at early times, of course one will require many terms to reach a given precision, compared to approximation at late times. However, provided one takes enough terms, the series will sum to give the required level of approximation. Thus we see, at least for flat sliced FLRW, that the late time expansion actually works rather better than expected in the sense that it converges for all the history of the universe, back to the epoch where other matter forms become relevant. Adding spatial curvature to the FLRW model appears not to change this picture, although we have not proved this, but have observed in examples that convergence extends back to the big bang, even for large curvature components.

We now make a brief aside on radiation which we have not included in our model. Radiation enters the late time expansion at order $y^4$. In principle we should include this when discussing terms in the expansion of order $y^4$ or higher. However, we note that since matter-radiation equality is at such an early epoch, $z \sim 3000$, the numerical size of the terms controlling radiation in the expansion must be \emph{unnaturally} small\footnote{This unnaturalness is, of course, a rephrasing of the usual small cosmological constant unnaturalness.}, of order $\sim 10^{-12}$, compared to the size of the terms controlling matter (such as $\rho_0$ in the exact FLRW solution above). Thus whilst radiation in principle enters the expansion at order $y^4$, the size of the contributions would be very suppressed compared to the size of terms for matter and curvature provided one is studying late times in the universe, i.e. much after matter-radiation equality.

We have arrived at the rather remarkable conclusion that provided inhomogeneity and anisotropy are small, then our late time expansion presumably converges all the way back to the beginning of the matter era. From the perspective of using the late time expansion to model observations this is very encouraging, as presumably taking sufficiently many terms, one can obtain accurate approximations to the true solution far enough back to include all current (and future) supernovae data, even though high redshift data fall outside the cosmological constant dominated epoch.

However, having found that our solution converges much further back than $\Lambda$-matter equality we then see that actually the fact that we have deformed about the late time de Sitter solution is somewhat unattractive, in the sense that we only describe homogeneous dust approximately. One might expect that if inhomogeneity and anisotropy were small one would require only a few leading terms to accurately model our universe. However looking back to the matter epoch our late time expansion will require many terms in order to accurately approximate FRLW. 

We may phrase this another way. Corrections in inhomogeneity and anisotropy enter at the same level as the homogeneous dust component. Whilst we have separated the dynamical scale of $\Lambda$ from those of  inhomogeneity and anisotropy, we have not managed to also separate out homogeneous dust. We may see this concretely exhibited in the solutions above for the expansion \eqref{eq:soln} as anisotropic terms (those tensors not proportional to $\delta_{ij}$), and inhomogeneous terms (those involving derivatives) enter the expansion in the same way as isotropic homogeneous spatial curvature contributions, which go as $\bar{R} \delta_{ij}$, and dust terms, going as $h \delta_{ij}$.

This situation may be simply improved using the fact that since we know the homogeneous isotropic solution is simply the FLRW solution, we may directly factor it out of the expansion, leaving a late time expansion that only consists of terms that explicitly involve inhomogeneity or anisotropy. We may say that instead of deforming about late time de Sitter, we instead rewrite our solution as a deformation about late time expanding $\Lambda$CDM FLRW. In the next section we proceed to implement this.

\section{\label{resummation} Improvement by resummation: A late time expansion about FLRW}

Consider a general expanding FLRW solution written using our above late time expansion. Then we have $g_{ij} = a^2(y) \Omega_{ij}$, where $\Omega_{ij}$ is the round sphere metric for a closed universe, the flat metric $\delta_{ij}$ for flat spatial slices, or a hyperbolic metric for an open universe. In terms of the characterizing data, $\bar{g}_{ij} = \Omega_{ij}$ with $\bar{R}_{ij} = 2 \, k \, \bar{g}_{ij}$, with $k > 0$, $k = 0$ or $k < 0$ for the closed, flat and open cases respectively. Let us take the constant $\rho_0$ to characterize the dust density relative to the cosmological constant and spatial curvature. From above we see the dust is characterized by the trace of $h_{ij}$. Since we are interested in a homogeneous and isotropic solution, then we see that $h_{ij}$ takes the form, $h_{ij} = - \rho_0 / (3 H)^2 \Omega_{ij}$, so that asymptotically $\rho \sim \rho_0 y^3 + \ldots$. Thus we find that the FLRW solution is given by the late time expansion,
\begin{eqnarray}
\label{eq:FRW}
g_{ij}(y) &=& \Omega_{ij} \left( 1 + \frac{k}{(2 H)^2}  y^2 - \frac{\rho_0}{(3 H)^2} y^3 + \ldots \right) \; .
\end{eqnarray}
By construction we obtain the general FLRW solution, with its homogeneous and isotropic spatial curvature and dust as a late time series expansion about local flat sliced de Sitter.

However whereas for inhomogeneous anisotropic solutions we have no analytic control over the solution beyond our late time expansion, for the homogeneous isotropic solution we know the solution, simply by solving the FLRW equations. Let us denote the full FLRW solution above as, $g_{ij}(y) = a^2(y;k,\rho_0) \Omega_{ij}$, so that the function $a(y;k,\rho_0)$ is the solution of the equations,
\begin{eqnarray}
\frac{\partial^2_y a}{a} - \frac{2}{y} \frac{\partial_y a}{a} + \frac{1}{2} \left( \frac{ \partial_y a }{a} \right)^2 + \frac{k}{2 H^2 a^2}  & = & 0 \nonumber \\
\frac{\partial^2_y a}{a} - \frac{1}{y} \frac{\partial_y a}{a} &=& -  \frac{\rho}{6 y^2 H^2} \qquad
\end{eqnarray}
with the appropriate asymptotic behaviour $\rho \sim \rho_0 y^3 + \ldots$. The expanding solution to these odes may be straightforwardly computed numerically, or given as a Frobenius series expansion analogous to \eqref{eq:dustexp} (expression \eqref{eq:dustexp} is without curvature) to very high order. Alternatively one may manipulate the formal solution written implicitly in terms of the integral,
\begin{eqnarray}\label{adef}
\int^{ \left( \frac{a(y; k , \rho_0) }{y} \right)^{3/2} }_0 \frac{dx}{\sqrt{ \tfrac{1}{3} \rho_0 - k \, x^{2/3} + H^2 \, x^2 } } = - \frac{3}{2 H} \log{\frac{y}{y_0}} \quad 
\end{eqnarray}
where $y = y_0$ gives the position of the big bang. In practice the numerical approach is probably the easiest method to obtain the function $a(y;k,\rho_0)$ to a desired precision in terms of the variables $y, k, \rho_0$. 

In our original discussion we presented our late time expansion as a power series solution for the metric $g_{ij}(x,y)$ in equation \eqref{eq:metric1}. As we discussed above, some terms in the expansion did not vanish in the homogeneous isotropic limit, namely those that describe the correction to de Sitter represented by the expanding FRLW solutions. However, given that we know the form of the FLRW solutions, we can factor out --- or `resum' ---  precisely these terms by taking the metric,
\begin{eqnarray}
\label{eq:metric2}
ds^2  =  \frac{1}{y^2} \left( - \frac{dy^2}{H^2}  + a^2(y; \frac{1}{6} \bar{R}, - \frac{1}{3 H^2} h) \hat{g}_{ij}(x,y) dx^i dx^j \right) \quad
\end{eqnarray}
and now performing a late time expansion on $\hat{g}_{ij}(x,y)$,
\begin{eqnarray}
\hat{g}_{ij}(y,x) & = & \bar{g}_{ij}(x) + y^2 b^{(0)}_{ij}(x) + y^3 \tilde{h}_{ij}(x)  \nonumber \\
&& \quad + y^4 b^{(1)}_{ij}(x) + y^5 b^{(2)}_{ij}(x) + \ldots
\end{eqnarray}
where we denote the anisotropic part of a tensor $T_{ij}$ (with respect to the 3-metric $\bar{g}_{ij}$) as,
\begin{eqnarray}
\tilde{T}_{ij} & \equiv & T_{ij} - \frac{1}{3} \bar{g}_{ij} T \; .
\end{eqnarray}
Again, the characterizing data is given by the same 3-metric $\bar{g}_{ij}$, and tensor $h_{ij}$ that lives on the 3-geometry defined by $\bar{g}_{ij}$, and we note that as before, the tensor $h_{ij}$ must satisfy the same condition,
\begin{eqnarray}
\bar{ \nabla}^j h_{ij} - \bar{\nabla}_i h = 0
\end{eqnarray}
and hence can be thought of as being defined by a general perturbation of the 3-geometry (subject only to the trace inequality for physical dust density). Now the coefficients entering the expansion are found to be,
\begin{eqnarray}
 H^2 b^{(0)}_{ij} & = &   \tilde{R}_{ij} \nonumber \\
 H^4 b^{(1)}_{ij} & = & \frac{1}{48} \bar{\nabla}^2 \bar{R} \bar{g}_{ij}- \frac{1}{16} \bar{\nabla}_i\partial_j \bar{R} + \frac{1}{4} \bar{\nabla}^2 \tilde{R}_{ij} \nonumber \\
 && \quad -\frac{1}{6}\bar{R}\tilde{R}_{ij} - \frac{1}{2} \tilde{R}_{im} \tilde{R}^m_{\phantom{m}j} +\frac{1}{4} \tilde{R}_{mn}\tilde{R}^{mn} \bar{g}_{ij}\nonumber \\
H^2 b^{(2)}_{ij} & = & \frac{1}{30} \bar{\nabla}^2 h - \frac{1}{10} \bar{\nabla}_i \partial_j h + \frac{1}{10} \bar{\nabla}^2 \tilde{h}_{ij} \nonumber \\
&& \quad - \frac{7}{30} h \tilde{R}_{ij} - \frac{13}{120} \bar{R} \tilde{h}_{ij} +\frac{3}{5} \tilde{R}_{m(i} \tilde{h}_{j)}^{\phantom{j)}m}
\end{eqnarray} 
and we see that having resummed the homogeneous isotropic terms, we are left only with terms that vanish in the homogeneous isotropic limit. 

Using our resummed late time expansion we may \emph{exactly} describe expanding homogeneous isotropic $\Lambda$CDM cosmologies. Inhomogeneous anisotropic cosmologies asymptotic to expanding FLRW ones can be described as deformations about FLRW. Since the expansion is made in $\hat{g}_{ij}$, when we approximate a solution by truncating the expansion to a given order, we still maintain an exact treatment of the homogeneous isotropic dust and curvature. We have succeeded in separating the homogeneous isotropic dynamics from the inhomogeneous and anisotropic behaviour, allowing the latter to be systematically included as a correction. Since in our universe we might expect inhomogeneity and anisotropy to be rather small corrections, then one would expect fast convergence to the true solution taking few terms in $\hat{g}_{ij}$. Then a low order truncation of the expansion of $\hat{g}_{ij}$ should give very good accuracy back to epochs sufficiently early to include current and future supernovae data (say $z \sim 5$).

We now wish to emphasize an important point. Inhomogeneity and anisotropy can be treated in cosmological perturbation theory as an expansion in the amplitude of the metric and matter field perturbation about the homogenous isotropic FRLW solution. Our late time expansion method also provides an inhomogeneous anisotropic solution as a deformation about an expanding FRLW solution. The difference of course is that our expansion is non-perturbative in the amplitude of the metric and matter perturbations, but is instead arranged as a derivative expansion. 
Thus, as stated above, we cannot treat short wavelengths unlike cosmological perturbation theory, without going to high order (providing the series converges). However, we do treat deformations of FLRW non-perturbatively in the amplitude of the metric and matter field perturbations about FLRW. 

A simple example illustrating this is to take a boundary metric that is a squashed 3-sphere. Taking a round 3-sphere would yield spatial curvature, whose effect may be small compared to the dust and $\Lambda$ today if the curvature $y_{\text{now}}^2 \bar{R} / H^2 \ll 1$ where we are living at $y = y_{\text{now}}$. Obviously, choosing $h_{ij}$ appropriately this simply yields a closed FLRW solution. However, we may deform away from FRLW by taking instead a squashed 3-sphere. Again, the effect of the spatial curvature may be made small today compared to dust and $\Lambda$.
For weak squashing, the solution can be described by cosmological perturbation theory about a closed FLRW solution. However for strong squashing away from the round metric, the solution can no longer be described by perturbation theory, whereas it can be described by our expansion, and since the spatial curvature is small relative to dust and $\Lambda$, we would expect our series solution to be convergent back to early times in the matter era.

\section{Observations and the luminosity-distance redshift relation}

We now provide a physical example of the application of our late time solution. We compute the past light cone of a late time observer and the redshift of sources comoving with the dust fluid. We use this to compute the luminosity distance as a function of observer  position and observation direction, and demonstrate how such a measurement encodes the solution data $(\bar{g}_{ij}, h_{ij})$. Details of the calculations presented in this section can be found in the Appendix.\\

\subsection{Late time expansion about de Sitter}

Consider an observer and a monochromatic isotropic source of radiation both comoving with the dust. We choose coordinates such that the observer is situated at $(y,x^i)=(y_o,0)$ and the source at $(y,x^i)=(y_e,x_e^i)$. Now consider a null geodesic connecting the two, such that its spatial tangent at the observer position is given by $\bar{v}^i$, normalized such that $\bar{v}^i \bar{v}^j \bar{g}_{ij}(0)=1$.

\begin{figure*}[]
\includegraphics[width=0.6\textwidth]{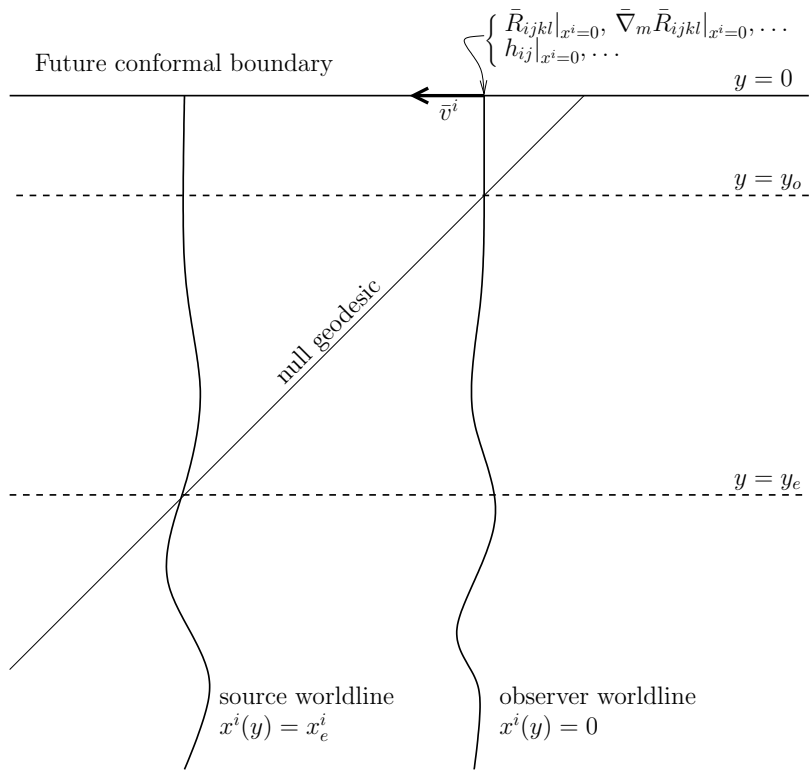}
\caption{The setup for computing the inhomogeneous luminosity distance}
\end{figure*}

The redshift, $Z$, of a photon emitted at $y_e$ and observed at $y_o$ may be calculated using the late time expansion in powers of $y$ as,
\begin{eqnarray}\label{bulk:redshift}
1+Z&=&\frac{y_e}{y_o} \Big(1-\frac{\left(y_e^2 - y_o^2\right)}{2H^2}\bar{V}^{mn}\bar{R}_{mn} \nonumber \\
&& \quad + \frac{1}{2} \big( - \frac{(y_e-y_o)^2(y_o +2y_e)}{3H^3} \bar{V}^{mn} \bar{v}^i \bar{\nabla}_i \bar{R}_{mn} \nonumber \\
&& \qquad + \left(y_o^3 - y_e^3\right)\bar{v}^i \bar{v}^j h_{ij}  \big) + O( y_e^4, y_e^3 y_o  \dots  y_o^4 ) \Big)\Bigg|_{x^i=0}
\end{eqnarray}
where $\bar{V}^{mn} \equiv \bar{v}^m \bar{v}^n - \frac{1}{4} \bar{g}^{mn}(0)$. The data $h_{ij}$, $\bar{g}_{ij}$, its curvature $\bar{R}_{ij}$ and higher derivatives are located at $(y,x^i)=(0,0)$, i.e. at the observers comoving coordinate position, but expressed at the future conformal boundary. For the luminosity distance, $D_L$, we obtain
\begin{widetext}
\begin{equation}\label{lumdist}
\begin{split}
D_L^2 = \frac{(1+Z)^2Z^2}{H^2}\Bigg(& 1 +  \frac{2(1+Z)}{H^2} y_o^2 \bar{V}^{mn}\bar{R}_{mn}  +\frac{Z(1+Z)}{H^3}y_o^3 \bar{V}^{mn}\bar{v}^i \bar{\nabla}_i \bar{R}_{mn}\\
&+  \frac{3(1+Z)(2+Z)}{2}y_o^3 h_{ij} \bar{v}^i \bar{v}^j + \frac{Z^2(2+Z)}{4}y_o^3 trh + O( y_o^4 )  \Bigg)\Bigg|_{x^i=0} \; .\end{split}
\end{equation}
\end{widetext}
The leading term gives the result for de Sitter with flat spatial sections. The first non-trivial term then arises from spatial curvature of the boundary metric. For homogeneous de Sitter with closed or open spatial sections this yields an isotropic contribution. However, we see that for our general inhomogeneous late time cosmology, the leading non-trivial contribution depends on the direction of observation, parametrized by the vector $\bar{v}^i$. At next order there is an isotropic contribution due to the trace of $h_{ij}$ from the dust fluid, an anisotropic contribution from $h_{ij}$ which we can interpret as being due to gravity waves, and further contributions due to inhomogeneity of the boundary metric which involve derivatives of the boundary curvature.

We note that our result here can be interpreted as an analog to the luminosity distance-redshift relation computed by Kristian and Sachs \cite{Kristian:1965sz}, where our result above is given for a Frobenius expansion about $y = 0$ (our late time expansion), and the result in \cite{Kristian:1965sz} can be thought of in terms of the local data at the position of the observer, which in turn can be thought of as a Taylor series expansion about some Cauchy surface containing the observers position. Consequently the calculations of \cite{Kristian:1965sz} amount to an expansion in small redshifts. Here, our results are valid to all orders in the redshift. As a check of our luminosity distance calculation \eqref{lumdist} in the low redshift regime, we have computed the luminosity distance using the geometrical optics method of \cite{Kristian:1965sz} for our late time solutions. We find precise agreement to the order of approximation in redshift considered in \cite{Kristian:1965sz}. As a further check we have verified that our expressions yield the correct isotropic and homogeneous results in the isotropic and homogeneous limit.\\

\subsection{Resummation: Late time expansion about FLRW}
The redshift (\ref{bulk:redshift}) and luminosity distance (\ref{lumdist}) were obtained above using the late time expansion technique, where the expansion was performed about de Sitter space. We now use the resummation technique described in section \ref{resummation} to improve these results. The key point is that even when the homogeneous and isotropic limit is taken, both still evaluate to an infinite series. This is because we have performed a late time expansion about flat de Sitter space. However we know how to obtain the solution in this limit, it is just the FLRW solution in the presence of dust, curvature and $\Lambda$. We can therefore factorize our results into a part which gives the redshift and luminosity distance for the FLRW solution, and a correction written as a late time series expansion, that describes the effects of inhomogeneity and anisotropy, and vanishes in the homogeneous isotropic FLRW limit. 

The resummed redshift and luminosity distance as a function of redshift are given by,
\begin{widetext}
\begin{equation}\label{resum:redshift}
\begin{split}
1+Z= & f\left(y_o,y_e;\frac{\bar{R}}{6},- 3 H^2 h \right ) \Bigg(1-\frac{\left(y_e^2 - y_o^2\right)}{2H^2}\bar{v}^i \bar{v}^j \tilde{R}_{ij} \\
& + \frac{1}{2}\left(\left(y_o^3 - y_e^3\right)\bar{v}^i \bar{v}^j \tilde{h}_{ij} - \frac{(y_e-y_o)^2(y_o +2y_e)}{3H^3} \left(\bar{v}^m\bar{v}^n \bar{v}^i \bar{\nabla}_i \tilde{R}_{mn} + \frac{1}{12} \bar{v}^i \bar{\nabla}_i \bar{R} \right)\right)+  O( y_e^4, y_e^3 y_o  \dots  y_o^4 ) \Bigg)\Bigg|_{x^i=0}
\end{split}
\end{equation}
and,
\begin{equation}\label{resum:lumdist}
\begin{split}
D_L^2(Z) = g\left(Z; \frac{\bar{R}}{6},- 3 H^2 h \right)^2 \Bigg(& 1 +  \frac{2(1+Z)}{H^2} y_o^2 \bar{v}^i\bar{v}^j\tilde{R}_{ij}  +\frac{Z(1+Z)}{H^3}y_o^3 \left(\bar{v}^m\bar{v}^n \bar{v}^i \bar{\nabla}_i \tilde{R}_{mn} + \frac{1}{12} \bar{v}^i \bar{\nabla}_i \bar{R} \right)\\
&+  \frac{3(1+Z)(2+Z)}{2}y_o^3  \bar{v}^i \bar{v}^j \tilde{h}_{ij}+ O(y_o^4) \Bigg)\Bigg|_{x^i=0} \; ,
\end{split}
\end{equation}
\end{widetext}
where,
\begin{equation}\label{FLRWredshift}
f(y_o,y_e;k,\rho_0) \equiv \frac{y_e}{y_o} \frac{a(y_o; k ,\rho_0) }{a(y_e;k, \rho_0)} \; ,
\end{equation}
and where the function $a(y; k, \rho_0)$ may be obtained from the integral in (\ref{adef}), and we have defined,
\begin{eqnarray}
g(y_o, Z; k,\rho_0) & \equiv & (1+ Z) \frac{a(y_o; k, \rho_0)}{y_o}\,  r(Z; k, \rho_0),\\
r(y_o, Z; k, \rho_0) & \equiv & \frac{1}{\sqrt{k} }  \sin \left[\sqrt{ k } \int_{y_o}^{y^*(Z)} \frac{dy}{H a(y; k,\rho_0) } \right], \label{rdef}
\end{eqnarray}
where $y^*(Z)$ is given by
\begin{equation}
1+Z= \frac{y^*(Z)}{y_o} \frac{a(y_o; k ,\rho_0) }{a(y^*(Z);k, \rho_0)}.
\end{equation}
In order to perform the resummation, we have factored out the quantities $f(y_o,y_e;k,\rho_0)$ and $g(y_o, Z; k,\rho_0)$. The advantage of using this form of the redshift and luminosity distance is that the factor involving the late time expansion is trivial in the homogenous and isotropic limit. In order to compute the part which does not vanish in this limit we have only to determine the FLRW scale factor function $a(y_o ; k, \rho_0)$ and perform the integral (\ref{rdef}). For practical purposes, where we wish to determine the model data through fitting to observations, we may simply numerically evaluate these. This should be contrasted with the late time expansion about flat sliced de Sitter (\ref{bulk:redshift}), where instead one must calculate terms in the late time expansion even to describe the homogeneous isotropic limit, and truncating this expansion would lead to an unnecessary approximation.

It is an interesting question whether all of the characterizing data can in principle be determined through observation. For example, truncating the late time expansion to the order we have worked with above, it is easy to show that all components of $\bar{R}_{ij}$, $h_{ij}$ and $\nabla_{(i} \bar{R}_{jk)}$ can be determined by extracting the quantities $\bar{R}$, $h$ and $\bar{v}^i \bar{v}^j \tilde{R}_{ij}$, $\bar{v}^i \bar{v}^j \tilde{h}_{ij}$ and $\left(\bar{v}^m\bar{v}^n \bar{v}^i \bar{\nabla}_i \tilde{R}_{mn} + \frac{1}{12} \bar{v}^i \bar{\nabla}_i \bar{R} \right)$ from comparison with luminosity distance-redshift observations for suitably many observation directions $\bar{v}^i$. However, at this order we cannot determine the non-symmetric parts of $\nabla_i \bar{R}_{jk}$, such as $\nabla_{[i} \bar{R}_{j]k}$.
Whether these non-symmetric parts can be determined from higher orders in the expansion is a question we defer for later investigation. 

To conclude, by measuring the dependence of the luminosity distance on redshift and angle on the sky (giving $\bar{v}^i$) an observer comoving with the matter can extract information about the characterizing data that lives at the future conformal boundary. For example, working to first non-trivial order they can in principle determine all components of the boundary Ricci tensor (and hence Riemann tensor) at the boundary location that they will asymptote to. At next order they determine all of $h_{ij}$ at this location and some information about derivatives of the Riemann tensor there. With sufficiently good data going to higher orders allows an observer to extract information about higher derivatives of $\bar{g}_{ij}$ and $h_{ij}$.

\section{Averaging solutions}

We have used a dust fluid to model the CDM, which is expected to be of a particulate nature. Hence a notion of averaging has been introduced to arrive at a smooth fluid description. It is then natural to consider how the inhomogeneous solutions transform under averaging. Our geometric characterization of a solution in terms of a 3-geometry and its perturbation suggest that a canonical averaging scheme can be found by considering an averaging of 3-geometries. Ricci flow provides precisely the analogy of diffusion for geometry, and as such gives an averaging of geometry. Introducing a flow time $\tau$, we have the Ricci flow equation,
\begin{eqnarray}
\label{eq:ricciflow}
\frac{d}{d \tau} \bar{g}_{ij}(\tau) = - 2 \bar{R}_{ij}[ \bar{g}(\tau) ] ,
\end{eqnarray}
where we take $\bar{g}_{ij}(0)$ to be our original metric $\bar{g}_{ij}$. Flowing for some time $\tau$ sets the scale of smoothing.

Now consider a metric $\bar{g}_{ij}$ and its perturbation $\bar{g}_{ij} + \epsilon \delta \bar{g}_{ij}$. Then we may Ricci flow both, and the difference between them after a time $\tau$ will be governed by a perturbation $\epsilon \delta \bar{g}_{ij}(\tau)$ where, initially $\delta \bar{g}_{ij}(0) = \delta \bar{g}_{ij}$ and its flow is given by the linearization of the Ricci flow about $\bar{g}_{ij}(\tau)$,
\begin{eqnarray}
\label{eq:linearizedricci}
\frac{d}{d \tau} \delta \bar{g}_{ij} = \triangle_L \delta g_{ij}  - 2 \bar{\nabla}_{(i} u_{j)}
\end{eqnarray}
where $\triangle_L$ is the Lichnerowiz operator defined by,
\begin{eqnarray}
\triangle_L \delta g_{ij} \equiv \bar{\nabla}^2 \delta \bar{g}_{ij} + 2 \bar{R}_{i~j}^{~m~n} \delta \bar{g}_{mn} - 2 \bar{R}_{(i}^{~~m} \delta \bar{g}_{j)m} \; ,
\end{eqnarray}
and $u_i$ is defined earlier in \eqref{eq:uvec}. The flows are invariant under the scaling $\delta \bar{g}_{ij} \rightarrow \lambda \delta \bar{g}_{ij}$ for $\lambda \in \mathbb{R}^+$.
Thus we take simultaneously flowing \eqref{eq:ricciflow} and \eqref{eq:linearizedricci} to provide a canonical smoothing of the data characterizing our late time cosmology, $( \bar{g}_{ij}, [ \delta \bar{g}_{ij} ] )$. Starting with initial data $(\bar{g}_{ij}(0), \delta \bar{g}_{ij}(0))$ one should perform the simultaneous flows for some time $T$, yielding smoothed data $(\bar{g}_{ij}(T), \delta \bar{g}_{ij}(T))$. One should then perform a coordinate transformation to present this data in harmonic gauge $u_i = 0$, and then use this to give the solution data $(\bar{g}_{ij}, h_{ij})$. We emphasize that as Ricci flow is a geometric flow, one may start with data $(\bar{g}_{ij}(0), \delta \bar{g}_{ij}(0))$ in any coordinate system and it will always yield the same result for $(\bar{g}_{ij}, h_{ij})$ up to diffeomorphisms. We note that the averaging scheme we naturally arrive at appears somewhat similar to the Ricci flow averaging considered by Buchert and Carfora \cite{Buchert:2002ht,Buchert:2003}. 

Let us consider smoothing data close to a flat FLRW cosmology. For the flat FLRW solution $\bar{g}_{ij} = \delta_{ij}$ and $\delta \bar{g}_{ij} = \tilde{\rho} \delta_{ij}$ for a constant $\tilde{\rho}$. This is a fixed point of our flow equations above. Moreover it is a stable fixed point. Firstly flat space is locally stable under Ricci flow, the flow existing for infinite time with small deformations `diffusing' away. Secondly consider a general perturbation $\delta \bar{g}_{ij} = \tilde{\rho} \, \delta_{ij} + E_{ij}$ now for non-constant $\tilde{\rho}$ and traceless $E_{ij}$, such that we have chosen harmonic gauge $u_i = 0$, so $\bar{\nabla}^j E_{ij} = \frac{1}{2} \bar{\nabla}_i \tilde{\rho}$, and take $\tilde{\rho} \rightarrow \mathrm{constant }$ and  $E_{ij} \rightarrow 0$ as $|x| \rightarrow \infty$.

Then about flat space $\bar{g}_{ij} = \delta_{ij}$, the flow \eqref{eq:linearizedricci} in harmonic gauge takes the simple form,
\begin{eqnarray}
\frac{d}{d \tau} \tilde{\rho} = \bar{\nabla}^2 \tilde{\rho} \, , \quad \frac{d}{d \tau} E_{ij} = \bar{\nabla}^2 E_{ij}
\end{eqnarray}
and therefore reduces to diffusion in flat space, which is well defined for all flow times, and renders $\tilde{\rho}$ constant at late time, and $E_{ij} \rightarrow 0$. We note that this flow preserves the harmonic gauge $u_i = 0$ we started in. The physical constraint $h < 0$ reduces to the condition $\tilde{\rho} < 0$, which is indeed preserved  under diffusion.
Thus for our proposed smoothing flows the flat FLRW cosmology model is locally a stable attractor, as we would physically wish.

Since Ricci flow is a highly non-linear operation, one will not be able to flow arbitrary starting data for an infinite flow time, and singularities may develop along the flow. For example the closed FLRW model is not a fixed point of our smoothing flow since it requires $\bar{g}_{ij}$ to be the metric on a round sphere and this shrinks to zero size in a finite flow time. Note also that for our solutions to be physical we require the dust density to be positive, and in general under our above flows this may not remain positive for arbitrary lengths of flow.

\section{Conclusion}

Suppose one wishes to consider spatial curvature in a cosmological model as is often done.
As we emphasized in the introduction, at least in the context of inflation and presumably more generally, it is unnatural to introduce spatial curvature without considering inhomogeneity and anisotropy. Once one introduces spatial curvature without assuming homogeneity and isotropy, one is typically in a regime beyond that described by cosmological perturbation theory. For example, as we have discussed above, if the spatial curvature is introduced as a squashed 3-sphere, rather than a round 3-sphere, then for any appreciable squashing, cosmological perturbation theory will fail as the solution is not a small metric deformation of FLRW. 
Thus if one wishes to observe or constrain the magnitude of spatial curvature in our universe, one is naturally led to the more general problem of characterizing curvature, inhomogeneity and anisotropy non-linearly in the amplitude of the metric and matter field perturbations. It is this characterization, and construction of the associated cosmological solution that has been the purpose of this paper.

Using Starobinsky's late time expansion (analogous to techniques developed for `holographic renormalization' in AdS/CFT) together with a resummation of the expansion, we are able to find the general inhomogeneous anisotropic solution to cosmological constant and dust fluid matter that is locally asymptotic to expanding de Sitter. The terms in the expansion describe a systematic deformation about a homogeneous isotropic expanding FLRW solution, and are arranged as a derivative expansion in the data that characterizes the solution.  The solution is fully non-linear in the amplitude of the metric and matter field deformations from FLRW. The leading terms in the expansion describe deformations on horizon scales, and higher terms describe deformations on increasingly small scales. The `final' data lives on the conformal boundary of the asymptotic de Sitter future. It is given by the boundary metric $\bar{g}_{ij}$ and the tensor $h_{ij}$ that lives on that boundary geometry, and as we have discussed, this pair can be elegantly thought of as defining a 3-metric and a perturbation of that 3-metric (up to an inequality on the trace of the perturbation that must be satisfied to ensure positive dust density).
Truncating the expansion at some finite order then allows a controlled approximation to the true  inhomogeneous anisotropic solution.
 
Recent cosmological observations indicate that we are resident in a universe whose matter content primarily consists of a cosmological constant and cold dark matter. We have argued that ignoring radiation is justified provided we restrict attention to the recent matter era ($z \ll 10^3$). We expect that for weak inhomogeneity and isotropy, the series given by our late time expansion should converge for this recent matter era, and furthermore, the convergence will be fast. Thus truncating the series at some low finite order will provide a good and controlled approximation to the true solution. Hence we hope that the solution is more than simply a formal construction, and may actually allow us to interpret observations in light of our assumptions that the Einstein equations hold and that the late time matter content is $\Lambda$ and dust fluid, without assuming anything about the homogeneity or isotropy of our universe. Ideally current and future structure observations, such as standard candle supernovae observations, could allow us to characterize systematically and precisely the large scale inhomogeneity and anisotropy, measuring them or constraining their size.

Our calculation of the luminosity distance-redshift relation to leading non-trivial orders in principle allows one to compare to large scale structure data and directly determine the Riemann tensor of the boundary metric and the value of the $h_{ij}$ tensor at the position corresponding to the observer position. These give the first anisotropic corrections to the solution. It allows one to extract partial information on the derivative of that curvature, giving $\bar{\nabla}_{(i} \bar{R}_{jk)}$, but not the antisymmetric components. This represents the leading inhomogeneity in the solution. Working to higher order in the expansion, one would extract more information about the future boundary data $(\bar{g}_{ij}, h_{ij})$. It would be interesting to understand whether there are parts of the data that are inaccessible by the luminosity distance-redshift measurement, and if so whether they might be determined by other types of observation, and precisely how much data can  be extracted in principle \cite{Ellis}. 

The use of an effective perfect fluid description of the dark matter already implies that an averaging has been performed. We consider averaging in the context of our inhomogeneous and anisotropic late time solutions. We argue that since the final data can be described in terms of a 3-geometry and its perturbation, it is natural to consider geometric flows, such as Ricci flows, to smooth the data. We consider Ricci flow of the boundary metric $\bar{g}_{ij}$ which then induces an associated flow on perturbations of that metric, and hence on the tensor $h_{ij}$. We confirm that using Ricci flow smoothing, a small perturbation away from flat sliced FLRW is smoothed towards the FLRW solution as we would wish.

\section*{Acknowledgements}

We would like to thank Carlo Contaldi, Fay Dowker, James Lucietti, and Dan Waldram for very enjoyable and useful discussions. TW would also like to greatly thank the wonderful teams of 7 South and West at Boston Children's Hospital where part of this work was completed, and in particular a special thanks is due to Mary Erbafina. TW is supported by an STFC advanced fellowship and Halliday award. BW is supported by an STFC studentship.

\appendix

\section{Luminosity distance-redshift relation}\label{appendix}

In order to construct the past light cone, we begin by considering null geodesics. We take a tangent to these $N^\mu$ so that $N^\mu N_\mu = 0$ and use the $y$ coordinate to parametrize the geodesic, so that $N^y = 1$, where the Greek indices $\mu = \{ y, i \}$. Then defining $N^i = n^i / H$ we find $\frac{d}{d y} n^i = N^\mu \partial_\mu n^i$ is given as,
\begin{eqnarray}\label{ap:geoeq}
\frac{d}{d y} n^i =  - \frac{1}{H} \Gamma_{mn}^{~~i} n^m n^n - \dot{g}_{mn} n^n \left( g^{im} - \frac{1}{2} n^i n^m \right)
\end{eqnarray}
which preserves the condition $n^i n^j g_{ij} = 1$ along the null geodesic. The null curve is parametrized as $x^i(y)$ so that $n^i = H d x^i(y) / d y$. In the usual way, one then obtains a second order o.d.e. for $x^i(y)$, which may be solved given some initial data constrained by $n^i n^j g_{ij} = 1$. Our late time solution \eqref{eq:soln} allows us to compute $g_{ij}$, $\dot{g}_{ij}$ and $\Gamma_{ij}^{~~k}$ in a $y$ expansion, and hence we may construct a solution to the null geodesic also in such a $y$ expansion. We now do this to the order at which the dust contributes in the $y$ expansion.

Consider a source, comoving with the dust, and emitting monochromatic radiation isotropically. We choose coordinates such that the source is located at $(y,x^i)=(y_e,x_e^i)$.
Consider a null geodesic which passes through this emitter point and reaches the future conformal boundary at $(y,x^i)=(0,\bar{x}^i)$. For calculational convenience we choose to write the spatial variation of the data $\bar{g}_{ij}(x)$ and $h_{ij}(x)$ as a Taylor expansion about the emitter's comoving coordinate position in the following coordinate system,
\begin{equation}
\begin{split}
\bar{g}_{ij}(x) =& \delta_{ij}-\frac{1}{3}\bar{R}_{ikjl}\Big|_{x^i=x_e^i}( x^k-x_e^k ) (x^l-x_e^l) \nonumber \\
& - \frac{1}{6}\left(\bar{\nabla}_k \bar{R}_{iljm}\right)\Big|_{x^i=x_e^i} ( x^k-x_e^k ) (x^l-x_e^l)( x^m-x_e^m )\\
&+ \mathcal{O}\left( x-x_e\right)^4 \nonumber \\
h_{ij}(x) =& h_{ij}\Big|_{x^i=x_e^i}+ \mathcal{O}\left(x-x_e\right).
\end{split}
\end{equation}
This coordinate system will simplify the calculations that follow, although we emphasize that the results we will obtain are coordinate invariant. The coordinates can be thought of as an extension of Riemann normal coordinates to higher order, and is discussed in an analogous context (null geodesics in late time expansions for pure cosmological constant matter) in \cite{Gibbons}. We recall that in 3-dimensions the Riemann tensor is simply given by the Ricci tensor as \eqref{eq:Riem}.

Later we will re-express this data in terms of the data at the observer's spatial position on the future conformal boundary. We may obtain the corresponding expansion for $a_{ij}^{(0)}$  to second order in $x^i-x_e^i$,
\begin{eqnarray}
H^2 a_{ij}^{(0)} &=& \left(R_{ij}-\frac{\delta_{ij}}{4}R\right)\Bigg|_{x^i=x_e^i} \nonumber \\
&& \, + \nabla_k\left(R_{ij}-\frac{\delta_{ij}}{4}R\right)\Bigg|_{x^i=x_e^i}(x^k-x_e^k) \nonumber \\
&&\, + \mathcal{O}\left(x-x_e\right)^2.
\end{eqnarray}
With data specified at $x^i=x_e^i$ the solution will not only be an expansion in $y$ but also in $y_e$. We will consider that $y$ and $y_e$ are of the same order in the expansion, so $O(y) \sim O(y_e)$. We employ the following ansatz,
\begin{equation}
n^i = v^i + \frac{y^2}{2!}\left(A^{i}+y_e B^{i}\right) + \frac{y^3}{3!}C^{i} + \mathcal{O}\left(y^4, y^3 y_e \dots y_e^4 \right).
\end{equation}
The data parametrizing this geodesic is $v^i$, and is the spatial tangent to the geodesic at $y=0$ on the future conformal boundary. We make this choice of parametrization for ease of calculation; later we will re-express this data in terms of the spatial tangent at the observer position. Integrating this expression we obtain the comoving coordinate along the geodesic,
\begin{eqnarray}\label{xexp}
H x^i\left(y\right) &=& H x_e^i +  (y-y_e)v^i + \frac{y^3 - y_e^3}{3!} \left(A^{i}+y_e B^{i}\right)\nonumber\\
&& +\frac{y^4 - y_e^4}{4!}C^{i}+\mathcal{O}\left(y^5, y^4 y_e \dots y_e^5 \right),
\end{eqnarray}
where the constant of integration has been fixed by the condition $x^i(y_e)=x^i_e$. With the expansions for $\bar{g}_{ij},h_{ij}$ and $n^{i}$ the geodesic equation can be solved as an expansion in $y$ and $y_e$,
\begin{equation}
\begin{split}
A^{i} =& v^{i} a^{(0)}_{jk}\Big|_{x^i=x_e^i}\, v^j v^k - 2  \delta^{ij} a^{(0)}_{jk}\Big|_{x^i=x_e^i}\,v^k\\
B^{i} =& - v^{i} v^l \frac{\left(\nabla_l a^{(0)}_{jk}\right)\Big|_{x^i=x_e^i}}{H}\, v^j v^k + 2 \delta^{i j} v^l \frac{\left(\nabla_l a^{(0)}_{jk}\right)\Big|_{x^i=x_e^i}}{H}\,v^k\\
C^{i} = & \delta^{ij}\frac{\left(\nabla_j a^{(0)}_{kl}\right)\Big|_{x^i=x_e^i}}{H}\,v^kv^l + 2 v^{i} v^l \frac{\left(\nabla_l a^{(0)}_{jk}\right)\Big|_{x^i=x_e^i}}{H}\, v^j v^k \\
&- 6 \delta^{i j} v^l \frac{\left(\nabla_l a^{(0)}_{jk}\right)\Big|_{x^i=x_e^i}}{H}\,v^k + 3v^i h_{jk}\Big|_{x^i=x_e^i}\, v^j v^k \\
& -6 \delta^{ij} h_{jk}\Big|_{x^i=x_e^i}\, v^k.
\end{split}
\end{equation}
The condition $g_{ij}n^in^j=1$ implies the data, $v^i$, is normalized as $1 = v^i v^j g_{ij}(y=0) = v^i v^j \bar{g}_{ij}(\bar{x})$. This yields the condition $v^i v^j \delta_{ij} = 1 + O(y^4)$ to the order required for our calculation. The data $\bar{g}_{ij}$ and $h_{ij}$ is specified at $(y,x^i)=(0,x_e^i)$. 


The redshift, $Z$, of a photon emitted at $y_e$ and observed at $y_o$ may be calculated as,
\begin{eqnarray}
 1+ Z  =  \frac{y_e}{y_o} e^{ \frac{1}{2} \int^{y_o}_{y_e} dy \dot{g}_{ij} n^i n^j }
\end{eqnarray}
where $n^i$ solves the null geodesic equation above. By considering \eqref{ap:geoeq} we deduce,
\begin{eqnarray}
\label{ap:redshift}
 1+Z&=&\frac{y_e}{y_o}\Big(1-\frac{\left(y_e^2 - y_o^2\right)}{2 H^2}V^{mn}\bar{R}_{mn} \nonumber \\
 && + \frac{1}{2}\Big(\frac{(y_e-y_o)^2(y_e +2y_o)}{3 H^3} V^{mn} v^i \bar{\nabla}_i \bar{R}_{mn}\nonumber \\
 && \quad + \left(y_o^3 - y_e^3\right)v^i v^j h_{ij} \Big)+ \mathcal{O}\left(y_e^4, y_e^3 y_o \dots y_o^4 \right)\Big)\Bigg|_{x^i=x_e^i}
\end{eqnarray}
where $V^{mn} \equiv v^m v^n - \frac{1}{4} \delta^{mn}$. 

We now consider the luminosity distance. Consider the congruence of null geodesics emitted from the point $y = y_e$, $x^i = x_e^i$, generated by taking the directions $v'^i = v^i + \alpha \delta v_1^i + \beta \delta v_2^i$, where $\delta v^i_{1,2}$ are infinitesimal vectors and $0 \le \alpha, \beta \le 1$. $\delta v^i_{1,2}$ must be linearly independent of each other and $v^i$, and must obey $v_i \delta v^i_{1,2} = 0$. A comoving dust observer sees this congruence trace out an infinitesimal 2-area $\delta A$, calculated as, 
\begin{equation}\label{ap:deltaadef}
\delta A = \frac{1}{y^3} \sqrt{g} \epsilon_{ijk} \left(y n^i\right) \delta x_1^j \delta x_2^k,
\end{equation}
where $\delta x^i_{1,2}$ give the variations in the position of the null geodesic at time $y$ due to the direction deformations $\delta v^i_{1,2}$, keeping the source position, $x_e^i$, fixed. The factor of $y$ preceding $n^i$ accounts for the fact that $n^i$ is normalized with respect to $g_{ij}$ rather than the spatial metric, $\frac{g_{ij}}{y^2}$. 

We now calculate the variation in $x^i(y)$ due to variations of the geodesic data, $v^i$, keeping $x_e^i$ fixed,
\begin{eqnarray}\label{ap:deltax}
H \delta x^i_N & = & \Big((y-y_e) \delta^i_j + \frac{y^3 - y_e^3}{3!} \left(\frac{\delta A^{i}}{\delta v^j} + y_e \frac{\delta B^{i}}{\delta v^j}\right) \nonumber \\
&& +\frac{y^4 - y_e^4}{4!}\frac{\delta C^{i}}{\delta v^j}+\mathcal{O}\left(y^5, y^4 y_e \dots y_e^5 \right)\Big)\delta v^j_N \; .
\end{eqnarray}
In order to compute \eqref{ap:deltaadef} we make use of the identity,
\begin{equation}\label{ap:identity}
\epsilon_{ijk} \left(T^j_a \delta v^a_1 \delta v^k_2 + T^k_a \delta v^a_2 \delta v^j_1\right) = \epsilon_{pmn}\delta v^m_2 \delta v^n_1\left(\delta^p_i trT - T^p_i\right),
\end{equation}
furthermore we note that
\begin{equation}
\epsilon_{pmn}\delta v^m_2 \delta v^n_1 =  v_p \delta\alpha
\end{equation}
for some constant $\delta\alpha$. To the order at which dust enters the expansion we find,
\begin{widetext}
\begin{equation}\label{ap:deltaa}
\begin{split}
\frac{\delta A(y)}{\delta \alpha} = \frac{(y-y_e)^2}{y^2 H^2}\Bigg( & 1 -\frac{y_e(y+y_e)}{4 H^2}\bar{R} + \frac{(2 y+y_e)}{2 H^2}\bar{R}_{ij}v^iv^j\\
& + \frac{y_e(y^2+y y_e + y_e^2)}{2}h_{ij}v^iv^j +\frac{(y^3-y^2y_e-y y_e^2 -y_e^3)}{4}trh\\
& + \frac{y_e(y^2-2y y_e -y_e^2)}{3 H^3}v^k \bar{\nabla}_k \bar{R}_{ij}v^iv^j - \frac{y_e(y^2 - 2y y_e -2 y_e^2)}{12 H^3}v^k \bar{\nabla}_k \bar{R}+ \mathcal{O}\left(y^4, y^3 y_e \dots y_e^4 \right)\Bigg)\Bigg|_{x^i=x_e^i} \; .
\end{split}
\end{equation}
Integrating this infinitesimal area over all possible data for the null geodesic $v^i$ whilst keeping $y$ fixed defines the area of a closed 2-surface surrounding the emitter position, $A(y)$. Then the solid angle of the congruence in the rest frame of the emitter can be found from  $ \delta \Omega =4\pi \lim_{y \rightarrow y_e} \frac{\delta A(y)}{A(y)}$. Then the luminosity distance, $D_L$, seen by an observer at $y = y_o$, is defined by, $D_L^2  \equiv  (1+Z)^2 \frac{\delta A(y_o)}{\delta \Omega}$. In practise we compute $A(y)$ using angular variables, $v = \cos(\theta)\partial_1 + \sin(\theta) \cos(\phi) \partial_2+ \sin(\theta) \sin(\phi) \partial_3$ and choose $\delta v_1^i = \partial_\theta v^i$ and $\delta v_2^i = \partial_\phi v^i$, integrating over the full range of angles $0\leq \theta \leq \pi$, $0\leq \phi <2\pi$.  We find,
\begin{equation}\label{ap:lumdist1}
\begin{split}
D_L^2 = \frac{(1+Z)^2}{H^2}\left(1-\frac{y_e}{y_o}\right)^2\Bigg(&  1 + \frac{y_e(y_o-y_e)}{H^2}V^{mn}\bar{R}_{mn}+\frac{y_e(y_e-y_o)^2}{3H^3}V^{mn}v^k \bar{\nabla}_k \bar{R}_{mn}\\
&+\frac{y_e(y_o-y_e)(y_o+2y_e)}{2}h_{ij}v^iv^j+\frac{(y_e+y_o)(y_o-y_e)^2}{4}trh+ \mathcal{O}\left(y_e^4, y_e^3 y_o \dots y_o^4 \right) \Bigg)\Bigg|_{x^i=x_e^i}\\
\end{split}
\end{equation}
or by inverting (\ref{ap:redshift}), we can eliminate $y_e$ and obtain luminosity distance as a function of the redshift,
\begin{equation}\label{ap:lumdist2}
\begin{split}
D_L^2 = \frac{(1+Z)^2Z^2}{H^2}\Bigg(& 1 +  \frac{2(1+Z)}{H^2} y_o^2 V^{mn}\bar{R}_{mn}  - \frac{Z(1+Z)}{H^3}y_o^3 V^{mn}v^k \bar{\nabla}_k \bar{R}_{mn}\\
& + \frac{3(1+Z)(2+Z)}{2}y_o^3 h_{ij} v^i v^j + \frac{Z^2(2+Z)}{4}y_o^3 trh + \mathcal{O}\left(y_o^4 \right) \Bigg)\Bigg|_{x^i=x_e^i} \; .
\end{split}
\end{equation}

The expressions for redshift \eqref{ap:redshift} and luminosity distance \eqref{ap:lumdist2} are characterized by data at the emitter position and geodesic data $v^i$, the spatial tangent to the geodesic at the boundary, $y=0$. We wish to convert these expressions so that all data is at the observer position. We calculate, evaluated on the geodesic at the observer position,
\begin{equation}
\bar{R}_{ij}\Big|_{x_e} =\bar{R}_{ij}\Big|_{x^i=0}  - (y_o-y_e) \frac{v^k \bar{\nabla}_k \bar{R}_{ij}}{H}\Big|_{x^i=0} + \mathcal{O}\left(y_e^2, y_e y_o , y_o^2 \right),
\end{equation}
whilst data for the geodesic at $(y,x^i)=(0,\bar{x}^i)$ can be converted to the normalized direction vector at the observer position,
\begin{equation}
v^i = \bar{v}^i + \mathcal{O}(y^2)
\end{equation}
normalized at the observer position, $\bar{v}^i \bar{v}^j \bar{g}_{ij}(0)=1$. For the redshift we obtain,
\begin{eqnarray}
\label{ap:redshift2}
 1+Z&=&\frac{y_e}{y_o}\Big(1-\frac{\left(y_e^2 - y_o^2\right)}{2 H^2}\bar{V}^{mn}\bar{R}_{mn} \nonumber \\
 && \qquad \qquad- \frac{1}{2}\Big(\frac{(y_e-y_o)^2(2y_e +y_o)}{3 H^3} \bar{V}^{mn} \bar{v}^k \bar{\nabla}_k \bar{R}_{mn} - \left(y_o^3 - y_e^3\right)\bar{v}^i \bar{v}^j h_{ij} \Big)+\mathcal{O}\left(y_e^4, y_e^3 y_o \dots y_o^4 \right)\Big)\Bigg|_{x^i=0},
\end{eqnarray}
where $\bar{V}^{mn} \equiv \bar{v}^m \bar{v}^n - \frac{1}{4} \bar{g}^{mn}(0)$.
For the luminosity distance we find,
\begin{equation}\label{ap:lumdist4}
\begin{split}
D_L^2 = \frac{(1+Z)^2Z^2}{H^2}\Bigg(& 1 +  \frac{2(1+Z)}{H^2} y_o^2 \bar{V}^{mn}\bar{R}_{mn}  +\frac{Z(1+Z)}{H^3}y_o^3 \bar{V}^{mn}\bar{v}^k \bar{\nabla}_k \bar{R}_{mn}\\
&+  \frac{3(1+Z)(2+Z)}{2}y_o^3 h_{ij} \bar{v}^i \bar{v}^j + \frac{Z^2(2+Z)}{4}y_o^3 trh + \mathcal{O}\left( y_o^4 \right) \Bigg)\Bigg|_{x^i=0} \; .\end{split}
\end{equation}
\end{widetext}

\bibliographystyle{apsrev}
\bibliography{references}

\end{document}